\documentclass{article}

\usepackage{arxiv}

\usepackage[utf8]{inputenc} 
\usepackage[T1]{fontenc}    
\usepackage{hyperref}       
\usepackage{url}            
\usepackage{booktabs}       
\usepackage{amsfonts}       
\usepackage{nicefrac}       
\usepackage{microtype}      
\usepackage{lipsum}
\usepackage{amsmath}
\usepackage{graphicx}
\usepackage{multirow}
\graphicspath{{Images/}}

\title{\textbf{Toward the Prevention of Privacy Threats: How Can We Persuade Our Social Network Platform Users? }}

\author{
   Ramon Ruiz-Dolz \\
   Valencian Research Institute for Artificial Intelligence\\
   Universitat Politècnica de València\\
   \textit{raruidol@dsic.upv.es} \\
   \AND
   Jose Alemany \\
   Valencian Research Institute for Artificial Intelligence\\
   Universitat Politècnica de València\\
   \texttt{jalemany1@dsic.upv.es} \\
   \And
   Stella Heras \\
   Valencian Research Institute for Artificial Intelligence\\
   Universitat Politècnica de València\\
   \texttt{stehebar@upv.es} \\
   \And
   Ana García-Fornes \\
   Valencian Research Institute for Artificial Intelligence\\
   Universitat Politècnica de València\\
   \texttt{agarcia@dsic.upv.es} \\
}

\begin{document}
\maketitle
\begin{abstract} \normalsize
Complex decision-making problems such as the privacy policy selection when sharing content in online social networks can significantly benefit from artificial intelligence systems. With the use of Computational Argumentation, it is possible to persuade human users to modify their initial decisions to avoid potential privacy threats and violations. In this paper, we present a study performed over 186 teenage users aimed at analysing their behaviour when we try to persuade them to modify the publication of sensitive content in Online Social Networks (OSN) with different arguments. The results of the study revealed that the personality traits and the social interaction data (e.g., number of comments, friends, and likes) of our participants were significantly correlated with the persuasive power of the arguments. Therefore, these sets of features can be used to model OSN users, and to estimate the persuasive power of different arguments when used in human-computer interactions. The findings presented in this paper are helpful for personalising decision support systems aimed at educating and preventing privacy violations in OSNs using arguments.
\end{abstract}

\keywords{Persuasion \and argumentation \and privacy \and human-computer interaction \and social network platforms }

\vfill

\section{Introduction}

Deciding which privacy policy is the best when making a publication in an Online Social Network (OSN) is not an easy task for human users because it requires to take multiple factors into account (i.e., the potential receivers, the information to be shared, the users' preferences, etc.). In many situations, the information regarding those factors can be incomplete or unknown, as the reachability of the publication or other users' preferences. Another relevant feature that characterises online communication is that, once the content is published online, it can be downloaded and stored by anyone with access to it. Therefore, it is important to make sure that the content published does not cause any future privacy issues. Additionally, if more than one user appears in the publication, it is even easier to violate any privacy preference of the rest of the users involved, leading to privacy conflicts between users. The multi-party privacy conflicts \cite{such2016resolving} are a common type of privacy threats happening in OSNs. This problem combined with the great increase of users in OSNs, mostly teenagers who are initiating in their usage and have limited abilities for self-regulation and complex decision-making \cite{DBLP:journals/ijcci/QuayyumCJ21}, has raised the interest of privacy management assistance research. 

A natural way to approach the existing privacy management problem in OSNs is with the use of Computational Argumentation \cite{ruiz2020towards}. Computational Argumentation research investigates how the human argumentative reasoning process can be approached from a computational viewpoint. Using Computational Argumentation techniques, it is possible to establish an argument-based human-computer interaction. This approach can be seen as a direct improvement of recommendation technologies \cite{chesnevar2009empowering} since added to the recommendation, a justification (i.e., an argument) is also provided to the user. An effective way to avoid and reduce the number of potential privacy threats (i.e., disclosure of sensitive information) is to persuade the author to adapt the initial privacy configuration since it may be harmful to him/her or any of the other users involved. The best way to persuade the author is by making him/her understand the reasons why the privacy threat is happening with the use of arguments. Using different messages and warnings make possible to persuade OSN users to modify their initial decisions \cite{alemany2019enhancing}. However, the perceived persuasive power of these messages may vary from one message to another \cite{thomas2017adapting} or even from different representations or structures of these messages \cite{thomas2019argumessage}. In the OSN privacy domain, these persuasive messages may approach different privacy aspects. Based on the previous definition given in \cite{ruiz2019towards}, up to four different types of argument might be considered depending on the source from which they can be supported: \textit{Privacy}, \textit{Content}, \textit{Risk}, and \textit{Trust}. Furthermore, arguments can be represented and structured according to different reasoning patterns. Argumentation schemes group the most common patterns of human reasoning \cite{walton2015classification}.

In this paper, we study the persuasive power of different argumentation schemes and argument types when used to educate teenagers on privacy management in an OSN environment. In our study, we consider two different user models that help us to encode human behavioural data into a computational system: the personality and the social interaction data. Previous work such as \cite{wu2018personalizing} show how users' personality can be a key factor when directly interacting with them. Therefore, we have investigated any potential existing correlations between personality traits and arguments. Additionally, since obtaining those traits may not be possible in some social network environments and behaviour is usually influenced by personality \cite{golbeck2011predicting, doi:10.1177/21582440211032156}, a study of the correlation between the most common social interaction features with the persuasive power of arguments has also been performed. Therefore, the main contributions presented in this work are the following:

\begin{itemize}
    \item Quantify and measure the persuasive power of arguments used as a privacy threat prevention mechanism.
    \item Analyse the existing significant correlations between the persuasive power of arguments and the Big Five personality traits.
    \item Analyse the existing significant correlations between the persuasive power of arguments and thirteen different social interaction features.
\end{itemize}

All these key findings are contextualised in the OSN educational domain with teenager participants. This is one of the most important target populations when working on this domain since they are very active and easy to convince to share their personal information.

The rest of the paper is structured as follows. Section \ref{sec:hcis-related} reviews the most relevant work regarding privacy management and argumentation in the OSNs domain. Section \ref{sec:hcis-background} introduces the background of our research, proposes the research questions, and presents the design of the study carried out in this work. Section \ref{sec:hcis-results} describes the observed results and analyses their implications and interpretation. Finally, Section \ref{sec:hcis-conclusions} summarises the most important conclusions reached at the end of this paper and the future research directions. 

\section{Related Work}
\label{sec:hcis-related}

Multiple approaches have been considered in the literature regarding the problem of finding optimal privacy policies in OSNs aimed at avoiding potential privacy violations \cite{DBLP:journals/csur/AlemanyVG23}. A collaborative privacy preserving tool proposed in \cite{DBLP:journals/sap/RussoBB21}. This system allows to provide recommendations to users that do not endanger their privacy. In \cite{DBLP:journals/ijicc/AltalbeK22} the authors propose an algorithm for predicting and preventing privacy violations in OSNs. This system detects a potential privacy violation and warns the involved users to prevent further damage to the parties involved. A different approach to the privacy problem was recently introduced in \cite{DBLP:journals/osnm/VolochGG21}, where the authors propose an algorithm that combines user features such as the age or gender with the trust between users to determine the risk of sharing a publication in an OSN. Another collaborative approach to provide privacy recommendations to users is proposed in \cite{squicciarini2011cope}. The authors propose \textit{CoPE}, a collaborative privacy management system where each user can decide a specific privacy configuration for each publication. The system decides the best policy considering the most voted configuration. Finally, some of the existing automated privacy management systems that rely on an internal negotiation process. For example, \textit{PriArg} \cite{kokciyan2017argumentation} is a multi-agent algorithm which has an underlying negotiation protocol to compute the best privacy configuration for a specific situation. In \textit{PriArg}, the negotiation is approached with argumentation. The agents represent real social network users that have an ontology with information from the network, the relationships between users and the content being published. Considering all these data, each agent can generate arguments to achieve a deal trying to satisfy the user privacy preferences. Images were also brought into consideration in \cite{DBLP:journals/aamas/KurtanY21}, where an autonomous agent uses the tags and image features to prevent privacy violations in OSNs. There are some common weak points in all these privacy management systems. All of them are focused on privacy conflicts where multiple users are involved in the same publication. But the case of a user choosing a dangerous configuration for itself is not considered. There is also an important limitation if we seek to provide the user with an explanation of why the configuration should be changed. None of the analysed privacy management tools gives the user a reasoned explanation nor tries to persuade him/her. A recent explainable approach was proposed in \cite{DBLP:conf/atal/MoscaS21, DBLP:journals/aamas/MoscaS22}, but it was focused exclusively on collective privacy violations.

When trying to reach an agreement, explaining our viewpoint, or trying to convince another person, it is very common to make use of arguments. An argument is defined as a set of propositions that can support the veracity of the main statement (the conclusion). Thus, with arguments, it is possible to provide a set of coherent reasons supporting some specific idea. Therefore, the use of Computational Argumentation can be seen as the natural way to approach a decision-making problem in which a human user must be persuaded. In \cite{hadoux2019comfort}, it is possible to observe the relevance of analysing the persuasive power of arguments and user preferences, when developing decision making assistance AI systems. Several works using argumentation in the OSNs domain can be identified in the literature. As described in \cite{Suyono_2021}, argumentation in OSNs can be very useful, such as enhancing dialogues or helping to structure user opinions. It is also possible to use Computational Argumentation techniques to model the dialogue between different users sharing their preferences in an OSN, and to persuade students to use specific learning objects in an educational environment \cite{heras2017using}. Therefore, as \cite{fogues2017sharing} proposes, argumentation seems the most coherent way to approach a persuasion problem framed in an educational context in OSNs. In \cite{kokciyan2017argumentation}, an argumentation protocol to define the best privacy policy when a multi-party privacy dispute is triggered is proposed. However, not many works in which all the topics of our research converge (i.e., privacy management, Computational Argumentation, and human user persuasion) have been identified. In addition to the main flaws identified before, the existing related work in argumentation in social networks is mainly focused on studying the multi-party privacy conflicts too \cite{kokciyan2017argumentation, DBLP:journals/aamas/MoscaS22}. As described in \cite{wang2011regretted}, it is very common to find users regretting their own publications in OSNs. Since we are focused on an educational domain, we need a system that considers not only privacy disputes between different users involved in the same publication but also potential self-privacy violations. When defining an argument, several parameters should be considered (e.g., the content, the reasoning pattern, the language, etc.) to maximise its persuasive power. The reasoning pattern of an argument is defined by the underlying logic of its elements. Argumentation schemes were conceived as common patterns on human reasoning. In \cite{walton2015classification}, up to sixty generally accepted argumentation schemes that can be found in common dialogues have been identified. Therefore, the use of argumentation schemes is a convenient way to define the reasoning patterns of the arguments of our study.  

Finally, persuasion plays a major role on the effectiveness of arguments when used in a dialogue. Different users may perceive arguments differently, so it is very important to be able to understand, and to adapt our arguments to each user if we want an effective human-computer interaction. In \cite{thomas2018argumessage}, an argumentative system to make users change their behaviour in the healthy eating domain is proposed. The persuasiveness evaluation of the semi-automatic generated arguments is described in \cite{thomas2019argumessage}. Furthermore, a study of the impact of personality, age, and gender on message type susceptibility \cite{ciocarlan2019actual} has also been done. Considering these works altogether, it is possible to infer relations between elements like personality and effectiveness of argumentation schemes. Finally, in \cite{DBLP:conf/umap/ruizumap22}, the authors explore the persuasive principles underlying some of the most common patterns of human argumentative reasoning (i.e., argumentation schemes). However, to the best of our knowledge, no one has directly analysed the persuasive power of arguments on teenagers, but behaviour may differ substantially between a teenager and an adult in the OSNs domain \cite{christofides2012hey}.

Therefore, with this paper, we put together these three research topics and present new results which will are helpful to push forward all the identified limitations on these topics: (i) with our arguments, we consider both self-disclosure and multi-party privacy conflicts; (ii) we approach the privacy management assistance problem from a more explainable and educational perspective; and (iii) we study teenager persuasion with arguments in OSNs, which has not been analysed in the literature yet. Our study results provide a new perspective on human (i.e., teenager) persuasion in the privacy management domain. We propose different, but related, user models based on two human aspects which we use to analyse the persuasive power of arguments: the personality and the social interactions. This way, it is possible to optimise the chosen argument by the privacy management assistance system for each specific user.

\section{Study Design}
\label{sec:hcis-background}

\subsection{Background}

Aimed at preventing privacy conflicts and minimising the number of privacy violations, an Argumentation Framework for Online Social Networks was proposed in \cite{ruiz2019towards}. It is defined as a tuple $\langle A,R,P,\tau_{p} \rangle$ where:
\begin{itemize}
    \item $A$ is a set of $n$ arguments $[$$\alpha_{1}$, \dots, $\alpha_{n}$$]$
    \item $R$ is the attack relation on A such as A$\times$A $\rightarrow$ R
    \item $P$ is the list of $e$ profiles involved in an argumentation process (i.e., a privacy dispute) $[$$p_{1}$, \dots, $p_{e}$$]$
    \item $\tau_{p}$ is a function $A\times P\rightarrow [0, \dots, 1]$ that determines the score of an argument $\alpha$ for a given profile $p$
\end{itemize}

A complete definition of all the parameters that define the Argumentation Framework for Online Social Networks is presented in \cite{ruiz2019towards}. As a solid motivation for the research conducted in this paper, it is important to mention that this framework models each user by their personality and their OSN usage statistics, which are the features that we analyse in this work. The personality of each user is represented with a 5-dimension vector modelled with the Big Five personality traits \cite{rothmann2003big}. These traits are \textit{Openness}, \textit{Conscientiousness}, \textit{Extraversion}, \textit{Agreeableness}, and \textit{Neuroticism}, that represent the five most significant aspects of human personality. The process of generating arguments is thoroughly explained in \cite{ramon2019automatic} and starts when a potential privacy violation is detected when publishing content in the social network. Then, the set of relevant information is gathered and retrieved from the OSN. Once all the arguments are generated, the system determines the set of \textit{acceptable} arguments based on the score function $\tau_p$. Finally, the system \textit{translates} the arguments in their computational shape into human readable text with the use of templates. The final step is the human-computer interaction. To interact with a human user, the argumentation system has the available the set of \textit{acceptable} arguments. However, the system needs to know which argument is more effective during the interaction process. The present work attempts to shed light on the persuasive power of argument types and argumentation schemes, to be able to define better dialogue strategies prioritising the most persuasive arguments.

\subsection{Research Questions}

The previously defined theoretical framework was proposed to be integrated into \textsc{Pesedia}, an educational social network \cite{alemany2020assisting}. However, deciding the dialogue strategies when interacting with human users is still a challenge. Therefore, we carry out this study to answer the following research questions that arise when designing this interaction:

\begin{enumerate}
    \item Which reasoning pattern (i.e., \textit{argumentation scheme}) is more persuasive for teenage OSN users?
    \item Which topic (i.e., \textit{argument type}) is more persuasive for teenage OSN users?
    \item How does the personality traits of teenage users influence the persuasive power of arguments?
    \item How does the online social interaction behaviour of teenage users influence the persuasive power of arguments?
\end{enumerate}

If it is possible to find any behavioural pattern regarding these questions, the arguments could be generated by the argumentation system following different strategies for each user depending on their personality traits or their social interaction behaviour.

\subsection{Measures and Instruments}

To answer the proposed research questions, we designed the following study based on three questionnaires and the social network usage. Questionnaires were used to retrieve the personality traits of the participants (Big Five personality test), the persuasive power of argumentation schemes (Questionnaire A), and the persuasive power of types of arguments (Questionnaire B). Participants also used the social network \textsc{Pesedia} \cite{alemany2020assisting} during one month from where we collected the online social interaction data. \textsc{Pesedia} is an educational OSN aimed at teaching its users the basic privacy competences in social networks. This social network provides a free environment like other OSNs (e.g., Facebook, Instagram, etc.). The chosen way to teach users is by gamification, with scores and a global ranking to reward the most active and participatory users. It is possible then, to nudge the users to do activities and participate in debates without forcing them \cite{alemany2019enhancing}. To find answers to the research questions proposed, this study has been carried out in \textsc{Pesedia} with teenage participants ranged from 12-15 years old. The study lasted one month, with the social network active and accessible 24/7 for participants. An ethics and law committee from the Universitat Politècnica de València reviewed and approved the study performed. Specifically, they reviewed that the social network \textsc{Pesedia} met the GDPR laws about users' privacy protection and management of their data.

Therefore, we measured their personality and online social interactions to model the participants of our study. A Big Five personality traits test aimed at measuring the personality traits (Openness, Conscientiousness, Extraversion, Agreeableness, and Neuroticism) of children and teenagers \cite{mackiewicz2016pictorial} has been used. Furthermore, we have also divided the participants into clusters based on their personality traits. Four major clusters have been recently identified in the literature: \textit{Average}, \textit{Self-centered}, \textit{Reserved}, and \textit{Role model} \cite{gerlach2018robust}. This clustering is proposed as a way to group samples with similar social perceptions and with similar expected behaviour. Our hypothesis to use these clusters in our study is that among similar characterised participants, it can be possible to observe stronger behavioural patterns, reducing the noise and leading to more solid findings. Thus, we have split our samples into four different personality-based groups to observe if those same clusters could be found in our population and if any behavioural pattern towards argument persuasion could be detected in each specific cluster. We ran the K-Means algorithm until its convergence to generate the mentioned clusters.

In some situations, it may not be possible to retrieve users' personality traits. Therefore, in our study, we have also considered the data from their social interaction behaviour in \textsc{Pesedia}. Thirteen different features representing participants' social interaction in the OSN have been used in our study to model \textsc{Pesedia} users as an alternative to their personality: number of friends (\#friends), number of status updates (\#status\_upd), number of likes (\#likes), number of shares (\#shares), number of comments (\#comments), number of private posts (\#ppprivate), number of public posts (\#pppublic), number of posts shared with friends (\#ppfriends), number of posts shared with collections of friends (\#ppcollections), number of uploaded photos (\#photos), number of posts deleted (\#deletes), the average length of text posts (avg\_textsize), and the time spent on the network (time\_spent). Previous work identified in the literature pointed out that these features could be closely related to user personality \cite{golbeck2011predicting, adali2012predicting, huang2019social, doi:10.1177/21582440211032156}. Therefore, these features represent an alternative dimension to personality from which it is possible to model OSN users.

\begin{figure}
    \centering
    \includegraphics[width=0.9\textwidth]{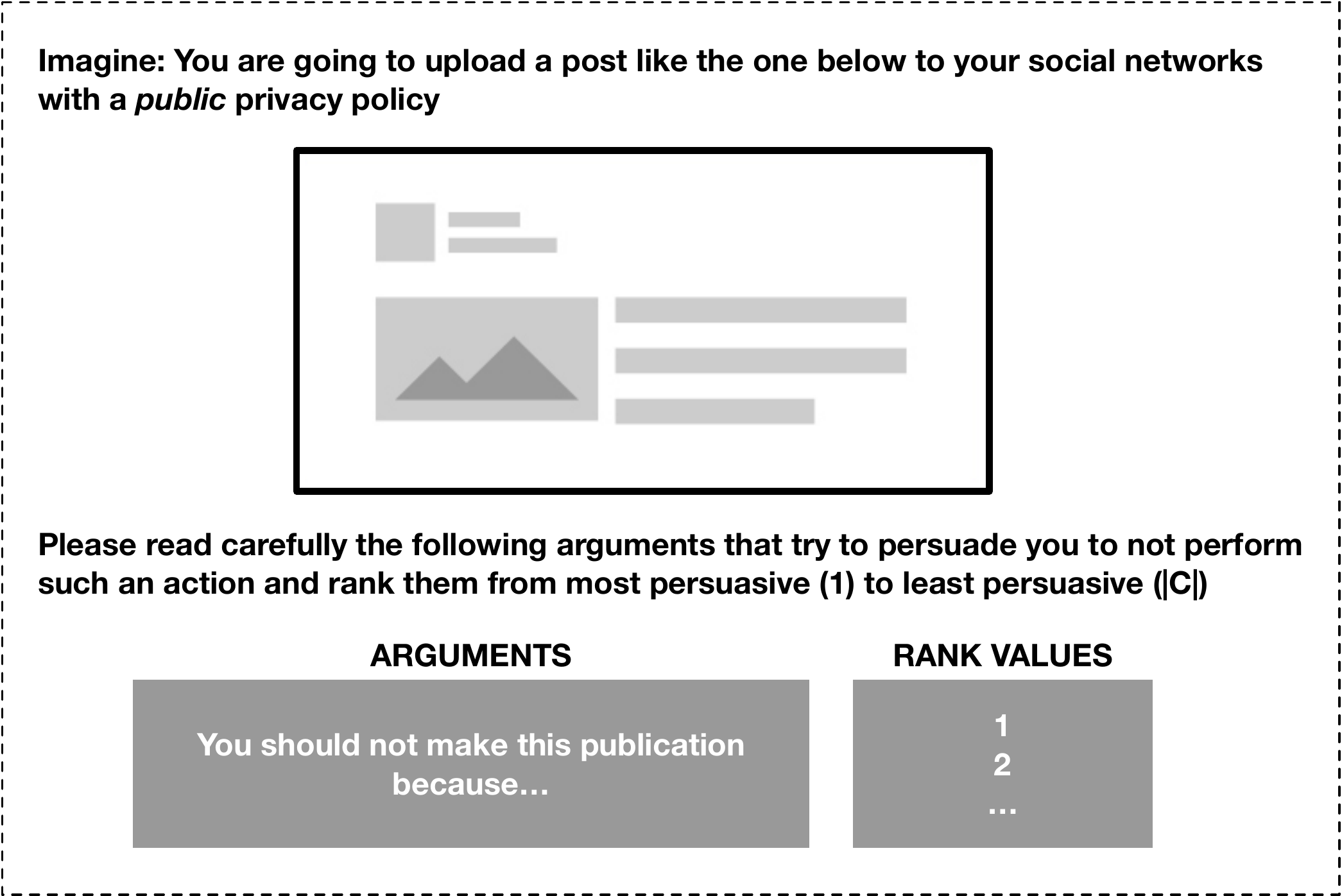}
    \caption{Template of the persuasive power questionnaires.}
    \label{fig:pptemplate}
\end{figure}

Finally, the persuasive power of arguments (for schemes and types) has been computed as the number of times an argument beats others. Our metric is based on \cite{dittrich2000analysis} work. Therefore, we define the persuasive power for an argument $\alpha_i$ as follows,

\begin{equation}
\label{eq1-hcis}
s(\alpha_i) = \frac{\sum_{j \in C} b_{ij}}{|P| \cdot (|C| -1)}
\end{equation}

where $b_{ij}$ refers to the number of times the argument $\alpha_i$ beats another argument $\alpha_j$ ($i,j \in C$, $i \neq j$). An argument $\alpha_i$ beats another argument $\alpha_j$ if it is considered more persuasive by our participants in the questionnaires. In our study, the classes $C$ are represented as argumentation schemes and types of arguments. Regarding the parameters $|P|$ and $|C|$, they represent the number of participants and the number of options inside a class and they are used to compute the maximum number of times an argument class can beat each other. The result is a 0-1 normalised value. We have used questionnaires to measure the persuasive power of arguments, different ones for schemes and types. There, participants were faced with the same situation (Figure \ref{fig:pptemplate}): they are going to make a publication in the network and they are told not to do it. The way of persuading the participant not to make the publication is with the use of arguments, so they had to rank these arguments, from the most persuasive argument ($1$) to the least persuasive one ($|C|$). Next, we describe how these questionnaires have been designed.

\subsubsection{Questionnaire A (Schemes)}

This questionnaire has been designed to capture the persuasive power of different argumentation schemes on a user (RQ1). We decided to consider the following five schemes in our study: \textit{Argument from Consequences} (AFCQ), \textit{Argument from Popular Practice} (AFPP), \textit{Argument from Popular Opinion} (AFPO), \textit{Argument from Expert Opinion} (AFEO), and \textit{Argument from Witness Testimony} (AFWT). With these schemes, it is possible to capture users behaviour when facing some of the most common reasoning patterns \cite{walton2008argumentation} used in social network privacy-related persuasive dialogues. Furthermore, we are able to analyse how practical reasoning and different source-based arguments are able to persuade teenager OSN users. By using these schemes, our goal was to see if teenagers were more concerned about recommendations based on the consequences of their actions, an expert opinion, similar user experiences, popular behaviours, or previously affected users.

\textit{Arguments from Consequences} show the participant the consequences of doing some specific action, sharing some content in our case. With this scheme, we can measure the importance each participant gives to the effect of their actions in the social network. \textit{Arguments from Popular Practice} try to persuade evidencing that there is a common popular practice among other similar people regarding some specific topic. In this case, with AFPP we can observe the importance that participants give to an argument based on their friends' activity. Similarly, \textit{Arguments from Popular Opinion} try to persuade with the use of a generally accepted opinion. Therefore, AFPO allows us to observe participants preferences towards the generally accepted opinion regarding their privacy. \textit{Arguments from Expert Opinion} base their reasoning pattern on some expert opinion regarding a specific topic. These argumentation schemes make it possible to observe users reliance in a privacy domain expert. Finally, \textit{Arguments from Witness Testimony} make the reasoning taking into account the experience of a person in the same knowledge position. With this scheme, it is possible to measure the trust that our participants give to someone with their similar expertise level in privacy management.

In this first questionnaire, the arguments that represent these five argumentation schemes in the OSN domain and that participants ranked by their perceived persuasive power are the following (You should not make this publication because...):

\begin{itemize}
    \item[$\bullet$] Making the publication could have bad consequences for your privacy (AFCQ)
    \item[$\bullet$] Most of your friends would not publish this content (AFPP)
    \item[$\bullet$] Everyone knows that publishing this is a mistake (AFPO)
    \item[$\bullet$] The monitors are experts in social networks and they believe that making publications of this type could be dangerous (AFEO)
    \item[$\bullet$] A user of the \textsc{Pesedia} network who has made similar publications considers that it can be dangerous (AFWT)
\end{itemize}

\subsubsection{Questionnaire B (Types)}

This questionnaire has been created to observe the persuasive power on our participants of the four different types of arguments considered by the argumentation framework (RQ2). These types are: \textit{Privacy}, \textit{Trust}, \textit{Risk}, and \textit{Content} arguments. \textit{Privacy} arguments are generated regarding each user privacy preferences towards the audience of his/her publications (i.e., private, friends, public, or friends collection). Therefore, \textit{Privacy} arguments will try to persuade the participants considering their privacy preferences and configuration. \textit{Trust} arguments are the ones generated taking friendships between users into account. This type of arguments will try to persuade the participant making him/her understand that other persons may be harmed if the content gets published. \textit{Risk} arguments consider the publication reachability in the network, computed as explained in \cite{alemany2018estimation,alemany2019metrics}. Then, a \textit{Risk} argument is generated if the scope of the publication exceeds the user expected audience. Finally, \textit{Content} arguments are generated regarding the own content of the publication. Six different types of content (i.e., location, medical, alcohol/drugs, personal information, family/association, offensive) \cite{caliskan2014privacy} are considered by our argumentation system. In this case, the degree of participants' persuasion may vary with the type of content included in the publication due to its sensitivity \cite{schomakers2019internet}. The arguments that participants ranked by their perceived persuasive power in this questionnaire and represent the four argument types are the following: (You should not make this publication because...)

\begin{itemize}
    \item[$\bullet$] you have chosen public privacy settings. (Privacy)
    \item[$\bullet$] some of the people who appear might get upset. (Trust)
    \item[$\bullet$] it could be read by strangers. (Risk)
    \item[$\circ$] you are revealing your location. (Content: Location)
    \item[$\circ$] you are giving out personal medical information. (Content: Medical)
    \item[$\circ$] others may think you consume alcohol/drugs. (Content: Alcohol/Drugs)
    \item[$\circ$] you are revealing personal data about yourself. (Content: Pers. Information)
    \item[$\circ$] you are revealing a friend's personal information. (Content: Fam./Assoc.)
    \item[$\circ$] you might offend some other user. (Content: Offensive)
\end{itemize}

where items represented as ($\bullet$) refers to Privacy, Trust, and Risk types of arguments and items represented as ($\circ$) refers to the different contents (Location, Medical, Alcohol/Drugs, Personal Information, Family/Association, and Offensive) of Content-type arguments. This questionnaire was done by participants as many times as different contents of Content-type of arguments are in order to avoid biases on users' perception of information sensitivity \cite{schomakers2019internet}.

\subsection{Procedure}

The study was carried out on the \textsc{Pesedia} social network where teenage users used it during one month. To prevent interferences, we included a registry controller (using a secret token) to avoid undesired registrations that could affect the security of the participants and the study. The questionnaires described above to measure participants' features were integrated in the own social network and they were progressively enabled in the on-site sessions. They were not required to complete them at any specific moment, but participants were motivated through gamification techniques. During the whole period of the study, the participants had fully access to the \textsc{Pesedia} social network to share their experiences and feelings.

We organised three on-site sessions of 90 min in equipped labs at the university to use as control points of the study. These three on-site sessions were distributed at three points in time: session 1, at the beginning of the one-month period; session 2, in the middle; and session 3, at the end. The aim of these sessions was to clarify any doubts that might arise among the participants about the functionality and features of the social network. Each session started with a brief explanation of the potential activities that they could do related to testing and understanding functionalities of the social network, and then participants had time to interact using the social network. In the first session, we introduced \textsc{Pesedia} to the participants and they signed up on the social network. Then, they had to complete basic activities that focused on customising their user profiles, setting up their general setting options, and building their friendship relations. Before finishing the first session, the personality test was made available for the participants to complete it. In the second session, we requested participants to complete the questionnaires about persuasive power (Questionnaires A and B). In Questionnaire A, participants ranked the five argumentation schemes in a decreasing persuasive ordering. In Questionnaire B, participants faced six different instances of the questionnaire considering one specific content category at a time. They ranked the four argument types in a decreasing persuasive ordering in each instance of the questionnaire. Arguments were displayed in a different order in each round to avoid the order effects. Finally, in the third (and last) session, we presented the participants with a summary regarding their behaviours and answers to the questionnaires to conclude the study.



\subsection{Participants}

A total of 218 teenagers participated in the study. From this total population, 215 participants completed the personality test and 212 completed both questionnaires A and B. We excluded the participants who did not complete all of the control sessions and the proposed questionnaires (29 participants) as well as the participants who decided not to participate (3 participants did not log into Pesedia). Finally, 186 participants completed the study\footnote{Contact the authors to get access to an anonymised version of the data gathered in this study.} (103 males, 83 females, $M_{age}$= 13.15, range: 12–15 years old). We included the participants in the experiment taking into account their age in order to have a sample of the teenage population (participants older than 12 years old). All of the selected participants were attending high school in different school centres of the Valencia area at the time of the experiment. In our study, we modelled our participants considering two different dimensions: the personality and the social interaction behaviour in the OSN. Furthermore, we investigated if stronger behavioural patterns could be identified when grouping our population by gender (i.e., male/female) and by personality clusters (i.e., \textit{Average}, \textit{Self-centered}, \textit{Reserved}, and \textit{Role model}). 



\begin{figure}
    \centering
    \includegraphics[width=0.9\textwidth]{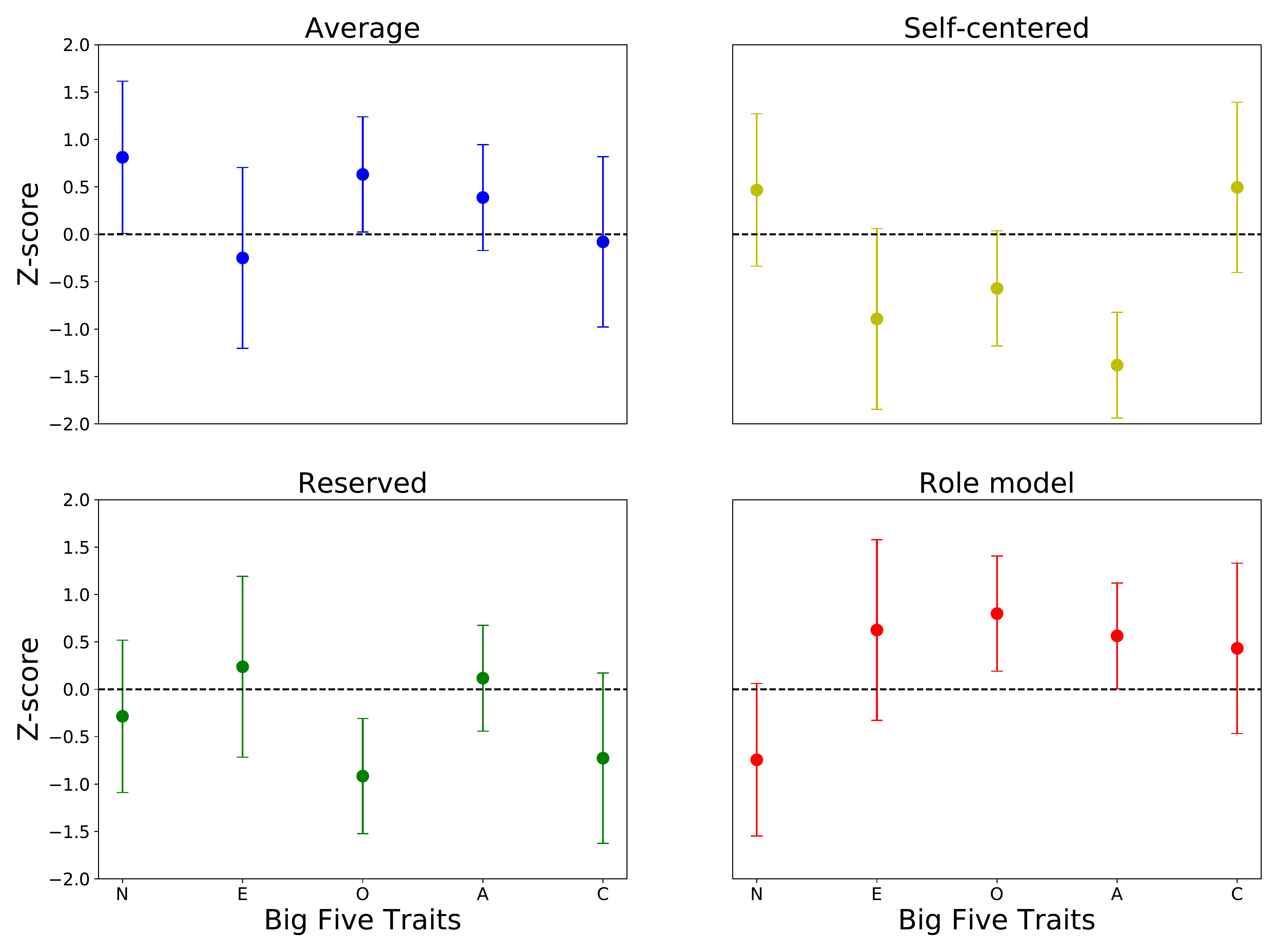}
    \caption{Personality clusters observed in our participants' data. ($\bullet$) Is the position of cluster centres represented as the average z-score of each cluster personality traits. The error bars represent the standard deviation of each trait in each cluster. The dotted lines represent global average values (Z=0) for each personality trait.}
    \label{fig:clustercomp}
    \vspace*{-0.3cm}
\end{figure}

The first modelling dimension considered in this research is the personality. We used the Big Five personality traits to represent the personality of our participants. From these five personality trait values, we grouped our participants into four different personality clusters. Those clusters had the following composition: the \textit{Average} cluster ($C_1$ = 44, 56.8\% males); the \textit{Self-centered} cluster ($C_2$ = 38, 68.4\% males); the \textit{Reserved} cluster ($C_3$ = 52, 48.1\% males); and the \textit{Role model} cluster ($C_4$ = 52, 63.5\% males). Figure \ref{fig:clustercomp} shows the Big Five personality traits distribution of the clusters found in our study. Each cluster is defined by the means of averages of each personality trait z-score. Therefore, it is possible to observe how depending on the cluster (i.e., \textit{Average}, \textit{Self-centered}, \textit{Reserved} and \textit{Role model}) the personality trait average z-scores of its members follow different distributions.

Comparing the clusters found in this work with the clusters proposed in \cite{gerlach2018robust}, it is possible to observe strong similarities between them. The Silhouette Coefficient (SC) \cite{rousseeuw1987silhouettes} of the computed clusters is 0.173 meaning that some clusters could be overlapping (SC $\approx$ 0) but the samples are not being miss-classified (SC$>$0). The \textit{Reserved} personality type is characterised by negative z-score values on Neuroticism and openness, while the rest of the traits (Extraversion, Agreeableness, and Conscientiousness) are slightly higher to 0. The \textit{Role model} personality type is characterised by negative z-score values on Neuroticism and positive z-score values for the rest of the traits. For both clusters, the personality traits of our participants followed the same distributions as \cite{gerlach2018robust} clusters. The \textit{Average} personality type is characterised by z-score values close to 0 for all personality traits. In our study, this cluster follows this trend with slightly higher z-score values on Neuroticism and Openness. Finally, the \textit{Self-centered} personality type is characterised by negative z-score values on all the personality traits except for the Extraversion trait. By comparing it with our cluster, we found some differences between those. However, we found a strong relationship with the original cluster in which \textit{Self-centered} was based, called \textit{Undercontrolled}, which was introduced in \cite{asendorpf2001carving} work. In \cite{gerlach2018robust}, the \textit{Undercontrolled} personality group is said to strongly influence the new proposed \textit{Self-centered} cluster. From the clusters observed in our population, we can support this statement. The \textit{Self-centered} cluster observed in this work has a positive Neuroticism z-score, similar to the original \textit{Undercontrolled} group. Furthermore, significant differences were observed regarding the Conscientiousness trait in our \textit{Reserved} and \textit{Self-centered} clusters. Studies have shown how Conscientiousness is the most variable trait with the age \cite{donnellan2008age}. Therefore, we think the observed differences were mainly due to the important age gap between the participants of both studies. 


Finally, the second dimension used to model OSN users in our analysis is their online social interaction behaviour. During our study, a total number of 2195 likes, 7650 comments, 1309 shares, 846 photos uploaded, and 7788 status updates (from them 761 were private, 769 were public, 5774 were disclosed to friends, and 484 were disclosed to specific lists of friends) were registered in the \textsc{Pesedia} database. The participants had a mean of 12 friendships and regretted 2761 actions made (which they undid/delete). 
The most common social interactions were comments and status updates. We observed an average of 41 comments and 42 status updates per user. 
It is also interesting to observe the high average number of deletes per user (i.e., 15), which represents a high number of regrets of the content published in the network. We also observed that in general, users preferred to share publications with friends only rather than publicly, privately or considering specific collections of friends. 

At the end of the study, we collected 186 different combinations of the Big Five personality traits and 18942 different OSN interactions. Furthermore, we also collected: 930 persuasive pairwise comparisons of argumentation schemes, one per participant (186) and argumentation schemes (5); and 4464 persuasive pairwise comparisons of argument types, one per participant (186), argument types (4), and contents variation of Content-type of arguments (6). 

\section{Results}
\label{sec:hcis-results}

Aimed at finding an answer to our research questions, we calculated the persuasive power of the five argumentation schemes and four argument types using the persuasive power equation (Equation \ref{eq1-hcis}). Furthermore, we did a correlation analysis between our user modelling features (i.e., personality and online social interaction features) and the previously calculated persuasive power values. In this section, we present the observed results after completing this process using the data gathered at the end of our study.

\subsection{Persuasive Power of Arguments}


From the results of the study, we have calculated the persuasive power of all the argumentation schemes and types considered in this work. Therefore, it is possible sort the five argumentation schemes taking into account their persuasive power as follows (RQ1): AFCQ $\succ$ AFEO $\succ$ AFWT $\succ$ AFPP $\succ$ AFPO. \textit{Argument from Consequences} seemed to be the most effective scheme for persuading our participants with a score of 0.61. Following, we have the \textit{Argument from Expert Opinion} scored with 0.53, \textit{Argument from Witness Testimony} with a score of 0.47, \textit{Argument from Popular Practice} with a score of 0.46 and finally, \textit{Argument from Popular Opinion} was the less persuasive scheme with a score of 0.43. These results mean that teenagers, in general, can be persuaded easier by showing the consequences of their actions or with recommendations made by experts rather than nudging them with recommendations made by someone similar to them or with popular trends or opinions. Table \ref{tab:confus-sch} represents the direct comparison of the persuasive ranking between pairs of argumentation schemes. For that purpose, we measure the number of times an argument is ranked in a higher position than another. We can notice how this value is higher when arguments with stronger persuasive power are compared with lower persuasive power arguments and vice-versa. 

\begin{table}
\caption{Pairwise rank comparative between argumentation schemes. This table represents the number of times an argumentation scheme (rows) beats another argumentation scheme (columns).}
 \label{tab:confus-sch}
\centering
\resizebox{0.6\textwidth}{!}{
\begin{tabular}{c c c c c c c}
\toprule
\textbf{} & \textbf{AFCQ} & \textbf{AFEO} & \textbf{AFWT} & \textbf{AFPP} & \textbf{AFPO} & \textbf{TOTAL}\\ \midrule
\textbf{AFCQ} & - & 106 & 114 & 112 & 123 & \textbf{455} \\
\textbf{AFEO} & 80 & - & 104 & 102 & 109 & \textbf{395} \\
\textbf{AFWT} & 72 & 82 & - & 96 & 99 & \textbf{349} \\
\textbf{AFPP} & 74 & 84 & 90 & - & 92 & \textbf{340} \\
\textbf{AFPO} & 63 & 77 & 87 & 94 & - & \textbf{321} \\
\bottomrule
\end{tabular}}

\end{table}

On the other hand, we sorted the argument types taking their persuasive power into account as follows (RQ2): \textit{Content} $\succ$ \textit{Privacy} $\succ$ \textit{Risk} $\succ$ \textit{Trust}. \textit{Content} arguments were the most persuasive with a score of 0.59. Following, we have \textit{Privacy} arguments with a score of 0.52, \textit{Risk} arguments with score of 0.47 and \textit{Trust} arguments with a score of 0.42. Meaning that teenagers are more concerned about sharing sensitive content rather than being read by unknown users or endangering other parties privacy. Similar to the previous analysis with argumentation schemes, Table \ref{tab:confus-typ} represents a direct comparison between the ranking position of every pair of argument types. Here, we also observe how arguments with a higher persuasive power score are ranked, in general, in a higher position than the rest. If we consider each round of the questionnaire B independently to analyse the effect of each content type on the persuasion of the argument, the following persuasive ordering is observed: Offensive $\succ$ Personal $\succ$ Family $\succ$ Medical $\succ$ Alcohol/Drugs $\succ$ Location. Therefore, although Content arguments were found as the most persuasive type of arguments, depending on which type of content was considered in each round, users' susceptibility was different. Our study revealed that teenagers are more concerned about sharing offensive content with a score of 0.64, closely followed by sharing personal information with a score of 0.62. The concern with these specific types of content matches the new trends in social networks of self-presentation \cite{chua2016follow}. The next most concerning types of content were family/association and medical content with scores of 0.59 and 0.58 respectively. Finally, revealing alcohol/drug consumption or location information, seemed to be the less relevant types of content for our participants with scores of 0.56 and 0.53 respectively.

\begin{table}
\caption{Pairwise rank comparative between argument types. This table represents the number of times an argument type (rows) beats another argument type (columns). }
 \label{tab:confus-typ}
\centering
\resizebox{0.6\textwidth}{!}{
\begin{tabular}{c c c c c c}
\toprule
\textbf{} & \textbf{Content} & \textbf{Privacy} & \textbf{Risk} & \textbf{Trust} & \textbf{TOTAL}\\ \midrule
\textbf{Content} & - & 623 & 659  & 681 & \textbf{1963} \\
\textbf{Privacy} & 493 & - & 605  & 634 & \textbf{1732} \\
\textbf{Risk} & 457 & 511 & - & 616 & \textbf{1584} \\
\textbf{Trust} & 435 & 482 & 500 & - & \textbf{1417} \\
\bottomrule
\end{tabular}}

\end{table}

\subsection{Personality Impact on Argument Persuasion}

To be able to adjust our argumentation system to increase the persuasive power of the arguments for our target population, we analysed the personality impact on the persuasive power (RQ3) of argumentation schemes and argument types. For this purpose, we have calculated the Spearman $\rho$ rank correlation between the persuasive power of arguments and the Big Five personality traits. In order to ease the interpretation and visualisation of the results, we have grouped correlations into three correlation-strength categories based in the ones proposed in \cite{corder2011nonparametric}. Weak correlations stand for correlation values between 0 and 0.2; we consider a Moderate correlation if its correlation value is between 0.2 and 0.6; finally, a Strong correlation stands for correlation values higher than 0.6.

The significant correlations found between argumentation schemes and argument types, with personality traits are represented in Table \ref{tab:conclusions-sch-b5} and Table \ref{tab:conclusions-typ-b5} respectively. As we can observe, different correlations have been found for different groups of users and associated with different personality traits. It is possible to observe how in some cases the personality correlates with the perceived persuasive power, which means that personality features could serve as predictors of persuasiveness when defining persuasive policies. We can also observe a smaller number of significant correlations between the argument types that the argumentation schemes. A possible cause for this pattern in our findings is that variations among different argumentation schemes have a greater impact on the perceived persuasiveness of an argument than the variation between argument types. Furthermore, studying specific groups of users categorised by descriptive features such as the gender or the personality allow to draw more informed conclusions than considering the whole heterogeneous group of users. Thus, personalisation is a key aspect to improve the effectiveness of the human-computer interactions of an argumentation system.

\begin{table}
\caption{Significant correlations of argumentation schemes persuasive power and personality traits. The significance is represented as: $\textbf{*}=p<0.05$, $\textbf{**}=p<0.01$. The correlation strength is represented as: Weak = $+/-$; Moderate = $++/--$; Strong = $+++/---$}
    \label{tab:conclusions-sch-b5}
\centering
\resizebox{0.8\textwidth}{!}{
\begin{tabular}{c c c c c c c}
\toprule
 & \textbf{Participants} & \textbf{O} & \textbf{C} & \textbf{E} & \textbf{A} & \textbf{N} \\ 
\midrule
 & \textbf{All} & - & - & $-$AFEO** & $-$AFPP** & - \\ 
\midrule
 \multirow{2}{*}{\textbf{Gender}} & \textbf{Male} & - & - & \begin{tabular}[c]{@{}c@{}}$-$AFEO*\\ +AFWT*\end{tabular} & \begin{tabular}[c]{@{}c@{}}$-$AFPP*\\ +AFWT*\end{tabular} & - \\
 & \textbf{Female} & - & - & - & - & +AFPP* \\ 
\midrule
 \multirow{4}{*}{\textbf{\begin{tabular}[c]{@{}c@{}}Personality\\ Cluster\end{tabular}}} & \textbf{Average}  & - & - & - & - & - \\
 & \textbf{Reserved} & - & - & $--$AFEO* & - & \begin{tabular}[c]{@{}c@{}}++AFPP**\\ $--$AFEO*\end{tabular} \\
 & \textbf{Self-centered} & - & $--$AFCQ* & - & - & - \\ 
 & \textbf{Role model}  & - & \begin{tabular}[c]{@{}c@{}}$-$AFPO*\\ ++AFEO**\end{tabular} & - & - & - \\
 \bottomrule
\end{tabular}}

\end{table}

\begin{table}
\caption{Significant correlations of argument types persuasive power and personality traits. The significance is represented as: $\textbf{*}=p<0.05$, $\textbf{**}=p<0.01$. The correlation strength is represented as: Weak = $+/-$; Moderate = $++/--$; Strong = $+++/---$}
 \label{tab:conclusions-typ-b5}
\centering
\resizebox{0.7\textwidth}{!}{
\begin{tabular}{c c c c c c c}
\toprule
 & \textbf{Participants} & \textbf{O} & \textbf{C} & \textbf{E}  & \textbf{A} & \textbf{N} \\ 
\midrule
 & \textbf{All} & - & - & $-$Privacy* & - & - \\ 
\midrule
 \multirow{2}{*}{\textbf{Gender}} & \textbf{Male} & - & - & - & - & - \\
 & \textbf{Female} & - & - & - & - & - \\ 
\midrule
 \multirow{4}{*}{\textbf{\begin{tabular}[c]{@{}c@{}}Personality\\ Cluster\end{tabular}}} & \textbf{Average} & - & - & - & - & - \\
 & \textbf{Reserved} & $-$Risk* & - & - & - & - \\
 & \textbf{Self-centered} & - & - & $++$Content* & - & - \\
 & \textbf{Role model} & - & - & - & - & - \\ 
\bottomrule
\end{tabular}}

\end{table}


\subsection{Social Interaction Impact on Argument Persuasion}

In some environments, obtaining the Big Five personality traits may not be possible. Therefore, in order to model our OSN users before analysing the persuasive power of arguments, we proposed an alternative to the personality based on the social interaction behaviour of our participants \cite{golbeck2011predicting, adali2012predicting, huang2019social, doi:10.1177/21582440211032156}. This way, we analysed if there existed any correlation between the persuasion of arguments towards each participant depending on their social interaction behaviour (RQ4). To measure the impact of these thirteen features on the persuasive power of arguments, we have calculated the Spearman $\rho$ rank correlation between them and the persuasive power of arguments. The interpretation of the correlation values is done the same way as the previous section. Furthermore, if personality traits are available, we have also considered making a complete analysis, taking into account personality clusters. This way, it is possible to combine the results of both analysis, thus observing even more useful correlations to define dialogue strategies.

The significant correlations found between argumentation schemes and argument types, with OSN interaction data have been represented in Table \ref{tab:conclusions-sch-social} and Table \ref{tab:conclusions-typ-social} respectively. As in the previous analysis, social interaction data has proven to be a good predictor of variations in perceived persuasion for different user models. Again, it was harder to find significant correlations when considering the argument types than the argumentation schemes. This pattern reinforces the hypothesis that argumentation schemes contain more persuasive information that argument types, and are aligned with the recent findings described in \cite{DBLP:conf/umap/ruizumap22}.


\begin{table}
\caption{Significant correlations of argumentation schemes persuasive power and social interaction data. The significance is represented as: $\textbf{*}=p<0.05$, $\textbf{**}=p<0.01$) The correlation strength is represented as: Weak = $+/-$; Moderate = $++/--$; Strong = $+++/---$}
\label{tab:conclusions-sch-social}
\centering
\resizebox{\textwidth}{!}{
\begin{tabular}{c c c c c c c c c c c}
\toprule
 & \textbf{Participants} & \textbf{\#friends} & \textbf{\#status\_upd} & \textbf{\#likes} 
 & \textbf{\#comments} & \textbf{\#ppprivate} & \textbf{\#pppublic} & \textbf{\#ppfriends} & \textbf{\#ppcollections} 
 & \textbf{avg\_textsize} 
 \\ 
\midrule
 & \textbf{All} & - & - & - & - & +AFPO* & - & - & $-$AFPP** & $-$AFPO* \\ 
\midrule
 \multirow{2}{*}{\textbf{Gender}} & \textbf{Male} & - & - & $--$AFPP* &  - & - & - & - & $--$AFPP** & - \\
 & \textbf{Female} & - & ++AFEO* & - & ++AFPO** & ++AFPO** & - & $--$AFCQ** & - \\ 
\midrule
 \multirow{4}{*}{\textbf{\begin{tabular}[c]{@{}c@{}}Personality\\ Cluster\end{tabular}}} & \textbf{Average} & ++AFEO* & - & - & - & $--$AFPP* & - & ++AFEO* & - & - \\
 
 & \textbf{Reserved} & ++AFCQ* & ++AFCQ* & - & - & - & $--$AFWT* & ++AFCQ* & - & - \\
 
 & \textbf{Self-centered} & - & - & - & - & $--$AFEO* & - & - & - & \begin{tabular}[c]{@{}c@{}}$--$AFCQ**\\ ++AFWT**\end{tabular} \\ 
 
 & \textbf{Role model}  & - & - & - & ++AFEO* & - & - & - & \begin{tabular}[c]{@{}c@{}}$--$AFCQ**\\ ++AFPO**\end{tabular} & ++AFCQ** \\

\bottomrule
\end{tabular}}

\end{table}

\begin{table}
\caption{Significant correlations of argument types persuasive power and social interaction data. The significance is represented as: $\textbf{*}=p<0.05$, $\textbf{**}=p<0.01$) The correlation strength is represented as: Weak = $+/-$; Moderate = $++/--$; Strong = $+++/---$}
\label{tab:conclusions-typ-social}
\centering
\resizebox{\textwidth}{!}{
\begin{tabular}{c c c c c c c}
\toprule
 & \textbf{Participants} 
 & \textbf{\#status\_upd} 
 & \textbf{\#comments} & \textbf{\#ppprivate} 
 & \textbf{\#ppcollections} 
 & \textbf{\#deletes} 
 \\ 
\midrule
 & \textbf{All} & - & - & - & - & -\\ 
\midrule
 \multirow{2}{*}{\textbf{Gender}} & \textbf{Male} & ++Trust* & - & - & ++Trust* & $--$Risk*\\
 & \textbf{Female} & - & - & - & - & - \\ 
\midrule
 \multirow{4}{*}{\textbf{\begin{tabular}[c]{@{}c@{}}Personality\\ Cluster\end{tabular}}} & \textbf{Average} & - & - & - & - & - \\
 & \textbf{Reserved} & - & - & - & - & - \\
 & \textbf{Self-centered} & - & - & $--$Risk* & - & - \\
 & \textbf{Role model} & ++Trust* & ++Trust* & - & - & - \\ 
\bottomrule
\end{tabular}}
\end{table}


\subsection{Interpretation of the Results}

This work sets the starting point to develop the human interaction part of argumentative educational systems to help with privacy management in OSNs. The findings observed in this paper reveal that personality traits and social interaction data are relevant user modelling features useful for estimating the perceived persuasive power of arguments by different user models. Therefore, these features represent a powerful way to model human users when approaching a problem of these specifications. These findings are consistent with recent research in similar topics \cite{hooshyar2022three}. In Table \ref{tab:example-models}, we present four different OSN user models considering the features proposed in this work. From the found correlations presented in the previous section, we estimate potential trends in the persuasive power of arguments for these users as depicted in Table \ref{tab:example-interpret}. We can observe how different user models may perceive arguments with a modified persuasive power. Thus, with the observed results, we can adapt the available argumentation schemes and argument types following different user tailored persuasive policies which are more effective than the one based on the general persuasive power of arguments.

\begin{table}
\caption{Four different user models. (-) represents an average value, ($\uparrow$) represents a value above the average and ($\downarrow$) represents a value below the average.}
\label{tab:example-models}
\centering
\resizebox{0.6\textwidth}{!}{
\begin{tabular}{c c c c c}
\toprule
 \textbf{Model Features} & \textbf{User 1} & \textbf{User 2} & \textbf{User 3} & \textbf{User 4} \\ 
 \midrule
 \textbf{Gender} & Male & Female & Female & Male \\ 
 \textbf{Cluster} & Self-centered & Role model & Reserved & Average \\ 
\midrule
\textbf{Openness} & - & - & $\downarrow$ & - \\ 
\textbf{Conscientiousness} & $\downarrow$ & $\uparrow$ & - & - \\ 
\textbf{Extraversion} & $\downarrow$ & - & - & $\uparrow$ \\ 
\textbf{Agreeableness} & - & - & - & $\uparrow$ \\ 
\textbf{Neuroticism} & $\uparrow$ & $\downarrow$ & $\uparrow$ & - \\ 
\midrule
\textbf{\#friends} & - & - & $\uparrow$ & $\downarrow$ \\ 
\textbf{\#status\_upd} & $\uparrow$ & - & $\downarrow$ & - \\ 
\textbf{\#likes} & $\downarrow$ & - & - & - \\ 
\textbf{\#shares} & - & - & - & - \\ 
\textbf{\#comments} & - & $\uparrow$ & - & - \\ 
\textbf{\#ppprivate} & $\downarrow$ & - & $\uparrow$ & - \\ 
\textbf{\#pppublic} & - & - & - & - \\ 
\textbf{\#ppfriends} & $\uparrow$ & - & $\uparrow$ & - \\ 
\textbf{\#ppcollections} & - & - & - & - \\ 
\textbf{\#deletes} & - & - & - & $\uparrow$ \\ 
\textbf{\#photos} & - & - & - & -\\ 
\textbf{avg\_textsize} & - & $\uparrow$ & - & $\downarrow$\\ 
\textbf{time\_spent} & - & - & - & -\\ 
\bottomrule
\end{tabular}}

\end{table}

\begin{table}
\caption{Persuasive power of argumentation schemes and argument types for four different users. (-) represents an unmodified value, ($\uparrow$) represents an increased persuasive power and ($\downarrow$) represents a decreased persuasive power.}
\label{tab:example-interpret}
\centering
\resizebox{0.5\textwidth}{!}{
\begin{tabular}{c c c c c}
\toprule
 \textbf{Persuasive Power} & \textbf{User 1} & \textbf{User 2} & \textbf{User 3} & \textbf{User 4} \\ 
 \midrule
 \textbf{AFCQ} & $\uparrow$ & $\uparrow$ & $\downarrow$ & -\\ 
\textbf{AFEO} & $\uparrow\uparrow$ & $\uparrow\uparrow$ & $\downarrow\downarrow$ & $\downarrow$ \\ 
\textbf{AFWT} & $\downarrow$ & - & - & $\uparrow$ \\ 
\textbf{AFPP} & - & $\downarrow$ & $\uparrow\uparrow$ & $\downarrow\downarrow$ \\ 
\textbf{AFPO} & $\downarrow$ & - & - & $\uparrow$ \\ 
\midrule
 \textbf{Content} & $\downarrow$ & - & - & -\\ 
\textbf{Privacy} & $\uparrow$ & - & - & $\downarrow$ \\ 
\textbf{Risk} & $\uparrow$ & - & $\uparrow$ & $\downarrow$ \\ 
\textbf{Trust} & $\uparrow$ & $\uparrow$ & - & - \\ 
\bottomrule
\end{tabular}}

\end{table}

Argumentation schemes have been previously investigated and classified by experts of many different disciplines such as spanning philosophy, communication studies, linguistics, computer science and psychology \cite{walton2015classification}. Thus, several clusters of schemes have been defined grouped according to their general category. The schemes we work with belong to the general categories of ``practical reasoning arguments" (AFCQ); and ``source-dependent arguments", concretely, to its subcategories of ``arguments from position to know" (AFEO and AFWT) and ``arguments from popular acceptance" (AFPO and AFPP). Recently, a relation between this classification and Cialdini's principles of persuasion has been established \cite{josekutty2019personalised, DBLP:conf/umap/ruizumap22}. Thus, the ``Consistency" principle of persuasion, by which people like to be consistent with the things they have previously said or done, relates to one's practical behaviour (AFCQ); the principle of ``Authority", by which people follow the lead of credible, knowledgeable experts, relates to source-based arguments (AFEO and AFWT); and the principle of ``Consensus", by which individuals are conformed to what the majority regards as acceptable, relates to arguments from popular acceptance (AFPO and AFPP). First, we can see how our findings detect a preference order ``Consistency" $>$ ``Authority" $>$ ``Consensus" for the persuasion principles in our social media domain. Second, although there is still no specific research that orders these persuasion principles by their persuasion power in the context of social media, similar research in the healthy eating domain concluded an order ``Authority" $>$ ``Consensus" and ``Consistency" (no significant difference between these two) and stated that persuasive power is highly influenced by the domain \cite{josekutty2019personalised}. Based on these mappings between argumentation schemes and persuasive principles, we can contextualise our findings within essential concepts of persuasive psychology research. Thus, any significant correlation detected between user descriptive features (i.e., personality and social interaction) and argumentation schemes can be also interpreted as a correlation between the user features and the three persuasive principles related to the five argumentation schemes analysed in our study. As an example, from our findings we can interpret that the ``Authority" (AFEO) principle has shown an increased persuasive power for \textit{Reserved} users with a low value on Extraversion. Although we have not been able to identify much prior research focused on this same purpose, the work presented in \cite{ciocarlan2019actual} sheds light on existing correlations between Big Five personality traits and Cialdini's persuasive principles. Albeit the populations of both studies differ substantially in age, some similarities can be identified among the significant correlations detected in both works. In Section \ref{sec:hcis-results}, we have identified the following correlations: a negative correlation between AFEO and Extraversion trait; a positive correlation between AFEO and Conscientiousness for our \textit{Role model} participants; and a positive correlation between AFPP and Neuroticism for our \textit{Female} and \textit{Reserved} participants. All these detected correlations have also been found in \cite{ciocarlan2019actual} work. However, the age gap and the significant differences between populations make it harder to compare the findings on both works and some of the found correlations remain unexplained.

\section{Conclusion and Future Work}
\label{sec:hcis-conclusions}

At the beginning of this work, we raised four different research questions aimed at having a better understanding of human persuasion in OSNs using arguments. With our findings, we have been able to answer the four research questions, and to have a better understanding of the persuasiveness of arguments when used with different user models. Personalisation plays a major role in effective human-computer interactions. In this paper, we have been able to observe that using representative user modelling features (i.e., personality and social interaction data) it is possible observe variations in the effectiveness of arguments. Therefore, the user models analysed in this work provide a solid basis for developing personalised argumentation systems aimed at educating and preventing privacy violations in an OSN environment. Given the nature of arguments and argumentation, the findings observed in our study lay the basis for developing powerful tools for education and decision-making assistance. 

As future work, we foresee to deepen the analysis presented in this paper by extending our study to an adult population, and to implement an argument-based persuasive algorithm able to generate personalised persuasive policies aimed at maximising the efficiency of human-computer interactions.

\bibliographystyle{plain}
\bibliography{references.bib}

\end{document}